\documentclass[a4paper,twocolumn,11pt,unpublished]{quantumarticle}
\pdfoutput=1
\usepackage[utf8]{inputenc}
\usepackage[english]{babel}
\usepackage[T1]{fontenc}
\usepackage{amsmath}
\usepackage{hyperref}

\usepackage{graphicx}
\usepackage{subfigure}
\usepackage{epstopdf}
\usepackage{graphicx}

\usepackage{amsfonts}
\usepackage{amssymb, color, bm}

\usepackage{currvita}

\newtheorem{proposition}{Proposition}

\usepackage{physics}

\usepackage{multicol}
\usepackage{lipsum}
\usepackage{mwe}

\DeclareMathOperator{\I}{\mathbb{I}}

\usepackage{tikz}
\usepackage{lipsum}

\begin{document}

\title{Dynamic Estimation Loss Control in Variational Quantum Sensing via Online Conformal Inference}

\author{Ivana Nikoloska}
%\affiliation{King's College London, Strand, London, WC2R 2LS, United Kingdom}
%\email{ivana.nikoloska@kcl.ac.uk}
\affiliation{Signal Processing Systems Group,  Department of Electrical Engineering, Eindhoven University of Technology, Eindhoven, 5612 AP, The Netherlands}
\email{i.nikoloska@tue.nl}
\author{Hamdi Joudeh}
\affiliation{Signal Processing Systems Group,  Department of Electrical Engineering, Eindhoven University of Technology, Eindhoven, 5612 AP, The Netherlands}
\author{Ruud van Sloun}
\affiliation{Signal Processing Systems Group,  Department of Electrical Engineering, Eindhoven University of Technology, Eindhoven, 5612 AP, The Netherlands}
\author{Osvaldo Simeone}
\affiliation{KCLIP lab, Center for Intelligent Information Processing Systems, Department of Engineering, King's College London, Strand, London, WC2R 2LS, United Kingdom}

\maketitle

\begin{abstract}
% Quantum sensing provides a route to overcoming practical limitations in classical sensing devices. As a result, it holds particular relevance in domains across science and engineering with sensitivity and resolution constraints. However, since quantum sensors are implemented using NISQ devices, they are challenged by noise effects as well as finite sampling rates. In this paper, we present an online control framework for variational quantum sensing (VQS), that can produce sensing estimates with error bars providing deterministic risk guarantees. Specifically, the proposed approach produces estimation sets with guaranteed long-term risk by leveraging online conformal risk control,  while supporting the online update of variational parameters.
% Experiments on a magnetometry task  confirm the practical benefits of the proposed dynamic estimation approach. 

%The framework is general and can be adapted towards arbitrary qubit architectures. End-to-end variational frameworks can thus underpin powerful design and analysis tools for realizing quantum advantage in practical, robust sensors.

Quantum sensing exploits non-classical effects to overcome limitations of classical sensors, with applications ranging from gravitational-wave detection to nanoscale imaging. However, practical quantum sensors built on noisy intermediate-scale quantum (NISQ) devices face significant noise and sampling constraints, and current variational quantum sensing (VQS) methods lack rigorous performance guarantees. This paper proposes an online control framework for VQS that dynamically updates the sensor’s variational parameters while providing deterministic error bars on the estimates. By leveraging online conformal inference techniques, the approach produces sequential estimation sets with a guaranteed long-term risk level. Experiments on a quantum magnetometry task confirm that the proposed  dynamic VQS approach maintains the required reliability (e.g., 90\% coverage) over time, while still yielding precise estimates. The results demonstrate the practical benefits of combining variational quantum algorithms with online conformal inference to achieve reliable quantum sensing on NISQ devices.
\end{abstract}

\section{Introduction}

\subsection{Context and Motivation}
Quantum measurements are ultimately governed by the laws of quantum mechanics. This  imposes fundamental limits on precision, but it also enables new sensing paradigms via non-classical effects such as coherence and entanglement. Quantum sensing leverages these effects to surpass classical sensor performance, with applications ranging from gravitational-wave observatories to biomedical imaging \cite{degen2017quantum,yu2022exposing,danilishin2012quantum,schwartz2019blueprint,budakian2024roadmap}. 

In practice, realizing these advantages is challenging: quantum sensors must be implemented on contemporary hardware that is subject to noise and size limitations. In particular, today’s noisy intermediate-scale quantum (NISQ) devices introduce decoherence and sampling errors that can degrade sensing accuracy. Designing near-optimal quantum measurement protocols for a given metrological task requires optimizing over high-dimensional Hilbert spaces, which is generally intractable with brute-force strategies.

\emph{Variational quantum algorithms} \cite{schuld2021machine,simeone2022introduction} have emerged as a promising approach to navigate the complexity of quantum optimization in the NISQ era. In the context of sensing, \emph{variational quantum sensing} (VQS) frameworks \cite{meyer2021variational,maclellan2024end,nolan2021machine,xiao2022parameter} use parameterised quantum circuits to prepare probe states and measurement operators that are adaptively optimised for the parameter estimation task. As illustrated in Fig. 1, a VQS system generates a probe state via a parameterised quantum circuit, exposes it to an unknown physical parameter through a quantum channel, and measures the output state to collect classical data. A classical algorithm then produces an estimate of the parameter. 

Prior works have demonstrated the potential of VQS \cite{meyer2021variational,maclellan2024end,nolan2021machine,xiao2022parameter} focusing on improving estimation precision. For example, reference \cite{meyer2021variational} introduced a variational toolbox for multi-parameter quantum estimation, and more recently the work \cite{maclellan2024end} realised an end-to-end variational sensor design on NISQ hardware. Machine learning techniques have been integrated into quantum sensing pipelines, such as neural-network-based Bayesian estimation \cite{nolan2021machine} and deep reinforcement learning for adaptive sensing policies \cite{xiao2022parameter}.

A common limitation of this prior art is the lack of formal guarantees on the estimation error. In fact, existing variational approaches follow a best-effort heuristic design and do not provide confidence intervals or risk bounds on the final estimates.

\subsection{Dynamic Estimation Loss Control}

In parallel, in classical machine learning and statistics, \emph{conformal inference} has emerged as a powerful tool for providing distribution-free confidence sets (or prediction intervals) that guarantee a user-specified coverage probability. In particular, significant progress has been made in online conformal inference, enabling uncertainty quantification under sequential data and potential distribution shifts. 

References \cite{gibbs2021adaptive,feldman2022achieving} introduced adaptive conformal inference methods that maintain valid coverage even as the data distribution evolves. Follow-up work \cite{angelopoulos2024online} proposed an online conformal prediction method with decaying step sizes to control long-term coverage  in an adaptive manner. Moreover, the work \cite{bhatnagar2023improved} improved online conformal prediction by incorporating strongly adaptive online learning, while the paper \cite{zecchinlocalized} extended the approach to accommodate localization guarantees. Applications of online conformal inference range from robotics \cite{dixit2023adaptive} and optimization \cite{deshpande2021calibration,kim2019attentive,zhang2024bayesian}, to wireless communication systems \cite{simeone2025conformal}. 

 These developments provide a principled way to guarantee reliability (in terms of coverage or error rate) for sequential predictions. However, they have not yet been applied in the quantum sensing domain, where one must contend with quantum measurement noise and the need to update quantum experiment settings on the fly.
 
Recognising the need to endow variational quantum sensors with formal performance guarantees, this work bridges VQS with online conformal inference, aiming to achieve deterministic estimation-loss control in a quantum metrology setting. By integrating an online conformal risk-control mechanism into the VQS loop, we obtain a framework that not only updates quantum sensor parameters adaptively but also outputs an estimation set (or “error bar”) at each time step with a guaranteed long-term risk level. For example, the method can ensure that the fraction of times the true parameter lies outside the provided estimation set is provably bounded by a specified value (e.g.,  10\%). 

\begin{figure*}
    \centering
    \includegraphics[width = 0.99\textwidth]{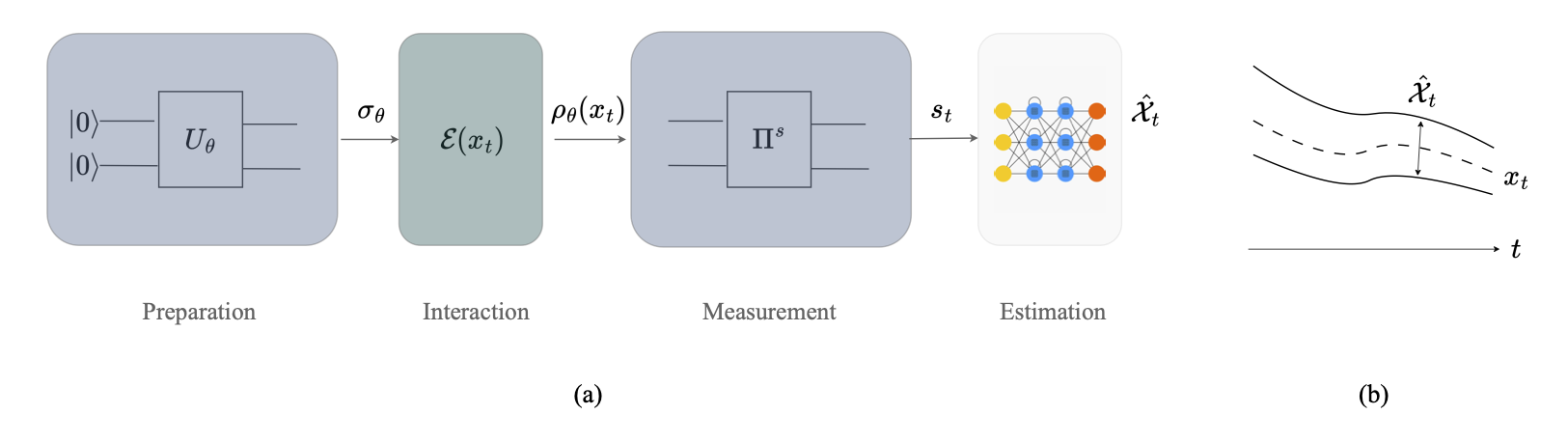}
    \caption{(a) Illustration of the variational  quantum sensing (VQS) system under study. A probe quantum state is generated by a $n$-qubit quantum circuit parametrised by the vector $\theta$. The probe state interacts with the parameter of interest $x_t$, and the perturbed state $\rho_{\theta}(x_t)$ is then measured with a given fixed POVM $\{\Pi^s\}$. The set of measurements $s$ is finally fed to a classical estimator, which produces an estimation set $\hat{\mathcal{X}}_t$. (b) The goal is to control the size of the estimation set  $\hat{\mathcal{X}}_t$ so as to meet a long-term deterministic reliability level.} 
    \label{fig:sensing_scheme_indirect} 
\end{figure*}

\subsection{Main Contributions}
In this paper, we propose a novel adaptive VQS protocol (illustrated in Fig. 1) that rigorously controls estimation loss while continually learning the optimal quantum sensor settings. The main contributions are as follows:

\noindent $\bullet$	\emph{Dynamic VQS with risk guarantees}: We develop an online control algorithm for VQS that integrates conformal inference into the quantum sensing cycle. At each discrete time step, the framework produces an estimation set for the unknown parameter -- e.g., a list if the parameter is discrete --  with a deterministic long-term coverage guarantee, while simultaneously updating the quantum circuit parameters to improve sensing performance. To this end, we assume access to feedback about the quality of the previous estimate, which can be used to adjust the confidence threshold on the fly, ensuring the cumulative risk (error rate) remains below a target $\alpha$. 

The proposed methodology may be applied to generic quantum sensing tasks during an initial training phase in which the sensed parameters are known by design. For example, in MRI the magnetic field is controlled by applying a known, periodic, sinusoidal modulation, with a known phase \cite{feiner1980nmr}. Training can be done in such controlled conditions and, assuming transferability, one could then use the same variational parameters/threshold when the phase is unknown and needs to be estimated.

More broadly, the method can be also deployed when quantum sensing is part of a control loop in which feedback is naturally obtained after acting on the sensing outcomes. For instance, an NV-center can be used to provide a measurement of the magnetic field in a quantum computer \cite{nobauer2015smooth, kumar2024high}, using which the magnetic field compensator can be adjusted. Since qubits might stay coherent or might decohere based on the compensation action, loss and feedback can then be accrued.

At a technical level, the proposed approach leverages online conformal inference techniques \cite{gibbs2021adaptive,feldman2022achieving,angelopoulos2024conformal}. To our knowledge, this is the first application of online conformal inference to quantum sensing, providing worst-case performance guarantees in a NISQ-based metrology task.

\noindent $\bullet$		\emph{Experimental validation}: We demonstrate the effectiveness of the proposed dynamic estimation scheme on a quantum magnetometry case study. The results show that our approach maintains the desired coverage -- more broadly, reliability -- over a long sequence of measurements, even in the presence of quantum noise and drift, outperforming baseline methods that lack online error control. In particular, the adaptive threshold mechanism successfully limits the empirical estimation loss to the predefined level (e.g., 0.1), while the variational updates yield tighter estimation sets (i.e., smaller uncertainty) compared to static or non-adaptive strategies. These findings highlight the practical benefit of integrating variational quantum algorithms with conformal inference to achieve robust and trustworthy quantum sensing.

The remainder of the paper is organized as follows. Section II formulates the problem of sequential quantum parameter estimation and defines the estimation loss metric to be controlled. Section III introduces the proposed dynamic VQS approach, including the conformal risk-control procedure and the online update rules for the quantum sensor. Section IV presents numerical experiments for a magnetometry task, detailing the setup, benchmarks, and results. Section V concludes the paper.

\section{Problem Formulation} \label{sec:qsens}

We consider the problem of sequentially estimating a target parameter $x_t \in \mathcal{X}$ determining a quantum channel $\mathcal{E} (x_t)$ over the discrete time index $t = 1,2, ...,T$. 

\subsection{Probes and Measurements}
As shown in Fig.~\ref{fig:sensing_scheme_indirect}, the channel is probed with an $n$-qubit quantum state $\sigma_{\theta}$, which is prepared by a quantum circuit parameterised by vector $\theta$. Probing results in a perturbed quantum state 
\begin{align}
    \rho_{\theta} (x_t) = \mathcal{E} (x_t) \sigma_{\theta},
\end{align}
where $\mathcal{E} (x_t) \sigma_{\theta}$ denotes the application of quantum channel $\mathcal{E} (x_t)$ to density matrix $\sigma_{\theta}$.

The perturbed state $\rho_{\theta} (x_t)$ is  measured via a fixed 
positive operator-valued measurement (POVM). The POVM is denoted as $\mathcal{M} = \{\Pi^s\}$, with $s$ being an integer in the set $[S] =\{0,1, ..., S-1\}$. The POVM may be, e.g., chosen as a standard local measurement in the computational basis. The resulting probability distribution $p_{\theta}(s \, | \, x_t)$ over measurement outcomes is 
%%%%%
%\begin{align}\label{probs}
%    p_{\theta, \mu}(s_m \, | \, \phi) = \text{Tr} \big[ \Pi_m(\mu) \rho(\theta, \phi) \big] 
%\end{align}
\begin{align}\label{probs}
    p_{\theta}(s \, | \, x_t) = \text{Tr} \big[ \Pi^s_{} \rho_{\theta}(x_t) \big] 
\end{align}
%%%%%%
for all $s \in [S]$.

\subsection{Loss and Feedback}

At each time step $t$ the estimator observes $L$ independent measurement outcomes, or shots, $\mathbf{s}_t = [s_{0,t}, ..., s_{L-1, t}]$ with $s_{l,t} \sim p_{\theta}(s \, | \, x_t)$ for all $l \in [L]$. Based on these observations, it aims at producing an estimation set $\hat{\mathcal{X}}_t \subseteq \mathcal{X}$ with the goal of providing a sufficiently informative estimate of the parameter $x_t$ over a given sequence of time steps $t=1,...,T$.

After having produced the estimation set $\hat{\mathcal{X}}_t$ and received the corresponding loss $L (x_t, \hat{\mathcal{X}}_t)$, the estimator also obtains  feedback about the true value of the parameter $x_t$. This may be provided by a human labeler or by a local measuring device. 
%For example, in magnetometry, this can be done during training by controlling the magnetic field by applying a known, periodic, sinusoidal modulation, with a phase that is known by the labeler \cite{feiner1980nmr}. {\color{red} can you provide some reference? {\color{blue} done} Also, can we  explain why we don't do this for sensing directly? i.e., why do we have to estimate xt first and then do this to get xt anyway?} {\color{blue} I just meant that training can be done in controlled conditions, when the magnetic field is known (like in MRI or NMR); then, assuming transferability, one could use the same variational parameters/threshold when the phase is indeed unknown and needs to be estimated (and there is no feedback). Still unsure if it is the best example.} 

The loss $L (x_t, \hat{\mathcal{X}}_t)$ of the set estimator $\hat{\mathcal{X}}_t$ is assumed to satisfy the following properties: 
\begin{itemize}
    \item \textit{Boundedness}: The loss function is upper bounded as $L (x, \hat{\mathcal{X}}) \leq L_{\text{max}}$ for some $L_{\text{max}} > 0$;
    \item \textit{Monotonicity}: For any two $\hat{\mathcal{X}}_t{'} \subseteq \hat{\mathcal{X}}_t{''} \subseteq \mathcal{X}$, we have the inequality
    \begin{align}
        L (x_t, \hat{\mathcal{X}}_t{'}) \geq L (x_t, \hat{\mathcal{X}}_t{''}),
    \end{align}
    for all $x_t \in \mathcal{X}.$
\end{itemize}
As examples, the loss can be defined as the \textit{coverage} loss
\begin{align}\label{hitt}
    L (x_t, \hat{\mathcal{X}}_t) = \I_{{x_t} \notin \hat{\mathcal{X}}_t} = \begin{cases}
        0, \,\,\, \text{if} \,\,\, x_t \in \hat{\mathcal{X}}_t \\
        1, \,\,\, \text{otherwise},
    \end{cases}
\end{align}
or as the minimum distance loss
\begin{align}
    L (x_t, \hat{\mathcal{X}}_t) = \underset{\hat{x}_t \in \hat{\mathcal{X}}}{\text{min}} \,\,\, || x_t - \hat{x}_t ||_p 
\end{align}
where $|| \cdot ||_p $ is the $L_p$ norm.

\subsection{Design Goal}

We are interested in controlling the average cumulative estimation loss over a time period of $T$ time instants, which is defined as
\begin{align}\label{barR}
    \bar{L}(T) = \frac{1}{T} \sum_{t=1}^{T} L (x_t, \hat{\mathcal{X}}_t).
\end{align}
%Specifically, we are interested in ensuring that
%\begin{align}\label{barRo}
%    \bar{L}(T) \leq \alpha + \mathcal{O} \left(\frac{1}{T}\right),
%\end{align}
%where $\mathcal{O} \left(\frac{1}{T}\right)$ ...
Specifically, the goal is to optimise the variational parameters of the probe $\theta$ and the mapping between observations $\mathbf{s}_t$ and set $\hat{\mathcal{X}}_t \subseteq \mathcal{X}$ under the constraint 
\begin{align}\label{barRo}
    \bar{L}(T) \leq \alpha + \mathcal{O} \left(\frac{1}{T}\right)
\end{align}
for some target average loss $\alpha$. By \eqref{barRo}, the
long-term loss $\bar{L}(T)$ is bounded by the target average loss $\alpha$ as $T \to \infty$ with a finite gap between $\bar{L}(T)$ and $\alpha$ decreasing at least as quickly as $\mathcal{O} (1/T)$.

\section{Dynamic Estimation via Online Conformal Inference}

This section introduces the proposed dynamic approach to the design of probes and estimator, together with the calibration of the  prediction set.

\subsection{Prediction Set}

At each time step $t$, based on the observation $\mathbf{s}_t$, the estimator assigns a score $C(\mathbf{s}_t, x)$ to each possible value $x\in \mathcal{X}$ of the target parameter. The score  $C(\mathbf{s}_t, x)$ is negatively oriented  -- i.e., the lower the better. With these scores, the estimator produces an estimation set by including in it all values $x$  whose scores are  below a threshold $\lambda_t$, i.e.,
\begin{align}\label{set_predict}
    \hat{\mathcal{X}}_t = \hat{\mathcal{X}}_t (\mathbf{s}_t, \lambda_t) = \{x \in \mathcal{X} : C(\mathbf{s}_t, x) \leq \lambda_t\}.
\end{align}

The scoring function $C(\mathbf{s}_t, x)$ can be in principle constructed from the likelihood  $p_{\theta}(\mathbf{s}_t \, | \, x)=\prod_{l=1}^L p_{\theta}(s_{l,t} \, | \, x)$ using  \eqref{probs}. For instance, a typical scoring function is the log-loss $C(\mathbf{s}_t, x)=-\sum_{l=1}^L \log p_{\theta}(s_{l,t} \, | \, x)$  \cite{angelopoulos2025theoretical}. However, evaluating the probabilities $p_{\theta}(s_{l,t} \, | \, x)$ for different values of $x$ would require having access to a classical description of the state $\rho_\theta(x)$, which may not be available. In fact, in practice, the state $\rho_\theta(x)$ may be affected by unknown quantum noise due to probe preparation; there may be errors associated with the channel model $\mathcal{E}(x)$; and the state may be excessively large to store and manipulate efficiently on a classical computer.

Therefore, we propose  replacing the true likelihood $p_{\theta}(\mathbf{s}_t \, | \, x)$ with a learned estimator  $p_{w}(x \, | \, \mathbf{s}_t)$ of the parameter $x$ given the observation $\mathbf{s}_t$. As indicated by the notation, the estimator  $p_{w}(x \, | \, \mathbf{s}_t)$ 
 is parameterised by a vector $w$. 

We specifically implement the estimator as a sequential model, such as  a recurrent neural network or a transformer, for the estimator $p_{w}(x \, | \, \mathbf{s}_t)$. A sequential model takes as input the sequence of measurement outcomes $\mathbf{s}_t=[s_{0,t},...,s_{L-1,t}]$, which are processed one by one along the index $l=0,...,L-1$, approximating  Bayesian inversion (see, e.g., \cite{muller2021transformers}).
 
 Given an estimator $p_{w}(x \, | \, \mathbf{s}_t)$, we consider the score\begin{align}\label{scoring_rule}
    C(\mathbf{s}_t, x) = - \log {p}_{w} (x \, | \, \mathbf{s}_t),
\end{align} which increases for values $x$ that are deemed less likely by the estimator.

As detailed below, the parameters $w$ of the estimator ${p}_{w} (x \, | \, \mathbf{s}_t)$ are  dynamically updated, together with the probe parameter $\theta$ and the threshold $\lambda_t$, with the received feedback over discrete  time $t$.

\subsection{Dynamic Calibration of the Prediction Threshold}

At the end of each time step $t$, after producing  the estimation set $\mathcal{X}_t$ and accruing the corresponding loss $L(x_t,\hat{\mathcal{X}}_t)$, the estimator receives feedback about the true value of the target parameter $x_t$. 
 Using this feedback, following the general framework of online conformal prediction \cite{angelopoulos2024online}, we propose to update the decision threshold $\lambda_{t}$ in (\ref{set_predict}) as
\begin{align}\label{lambda_update}
    \lambda_{t+1} = \lambda_{t} +  \eta_{t} \cdot (L (x_t, \hat{\mathcal{X}}_t) - \alpha),
\end{align}
where $\alpha$ denotes the user-defined target long-term loss in \eqref{barRo} and $ \eta_{t} > 0$ is a sequence of learning rates. The online update rule in \eqref{lambda_update} increases the threshold, thus making the estimation set $\hat{\mathcal{X}}_t$ in \eqref{set_predict} more inclusive, whenever the current loss exceeds the target level $\alpha$. Vice versa, the threshold is decreased if the current loss is below the target level $\alpha$, making it possible to reduce the size of the estimation set. 
%In the following, we provide a formal, long-term risk guarantee.
%The obtained set of predictions can be used to quantify the model uncertainty in a principled manner -- a wide set of predictions represents an uncertain model. In principle, the estimator can always reliably quantify uncertainty by including all possible predictions in the set; however, this is not useful to produce an actual estimate. To enable estimation in practice, the ansatz parameters $\theta, \mu$ are trained to minimise the size of the prediction set and thereby balance encoding uncertainty with the ability to produce an estimate. 
%\subsection{Long-term risk}
%In each epoch, the probes change the training data for the estimator. As a result, the learning problem is an instance of online learning with non i.i.d data. Thereby, when updating the base estimator in the $t$-th epoch, the distribution shifts, and in this case as established in \cite{angelopoulos2024online}, appropriately decaying step sizes $\eta_{t}$ can recover convergence. 

\subsection{Dynamic  Optimisation of Probe and Estimator}
At the end of each time step $t$, we  leverage the feedback $x_t$ not only to update the threshold (\ref{lambda_update}),  but also to optimise the probe parameters $\theta$ and the estimator parameters $w$ in order to improve precision and decrease the estimation uncertainty. In prior art, the optimisation of the probe parameters $\theta$ was typically done by minimising the Cramer-Rao lower bound, which measures the variance of an unbiased estimator in the  asymptotic regime of a large number of data points \cite{meyer2021variational, maclellan2024end}. In contrast, here we propose adopting the reliability property (\ref{barRo}) of the proposed threshold update (\ref{lambda_update}) (see Proposition 1 below) as an objective.
%to adopt as objective the time-averaged informativeness of the estimation set up to the current time $t$.

Specifically, we are ideally interested in minimising in an online fashion the time-averaged  cardinality $|\hat{\mathcal{X}}_t|$ of the estimation set $\hat{\mathcal{X}}_t$, i.e., \begin{equation}\label{eq:opt} \min_{\theta,w} \frac{1}{T} \sum_{t=1}^T |\hat{\mathcal{X}}_{\theta,w,t}|, \end{equation} where the notation $|\hat{\mathcal{X}}_{\theta,w,t}|$ makes it clear that the estimation set (\ref{set_predict}) with scoring function (\ref{scoring_rule})  depends on both the probe parameters $\theta$ and the estimator parameters $w$. This objective is motivated by the fact that a smaller predicted set satisfying the property (\ref{barRo}) is more informative, offering a smaller uncertainty while guaranteeing the desired level of loss $\alpha$. 

The objective function (\ref{eq:opt}) cannot be directly optimised due to the non-differentiability of the cardinality function $|\cdot | $. We address this problem by replacing the cardinality function with a smooth surrogate \cite{stutz2021learning,park2023few}. In particular, we use the approximation \begin{align}\label{soft_X}
    |\hat{\mathcal{X}}_{\theta,w,t}| \approx G_{\theta,w,t}=\sum_{x \in \mathcal{X}} \sigma\left(\frac{-(C(\mathbf{s}_t, x) - \lambda_t)}{\tau}\right),
\end{align}where $\sigma(\cdot)$ denotes the sigmoid function defined as
\begin{align}
\sigma(z) = \frac{1}{1+ \exp(-z)}
\end{align} and $\tau>0$ is a parameter. As the parameter $\tau$ tends to zero, the approximation  \eqref{soft_X} can be readily seen to be an equality \cite{stutz2021learning,park2023few}.

% the condition in \eqref{set_predict} can be written as
% \begin{align}\label{soft_set}
%     q (x, \theta) := 
% \end{align}
% where 
% and where $T > 0$ denotes a temperature hyperparameter. Thereby, the expression in \eqref{soft_set} can be interpreted as the probability that a particular $x \in \mathcal{X}$ is being included in the set $|\hat{\mathcal{X}}_t|$ at time $t$. As $T \rightarrow 0$, we have $q (x, \theta) \rightarrow 1$ for all $x \in \hat{\mathcal{X}}_t$, whereby

Using \eqref{soft_X}, the probe parameters $\theta$ and the estimator parameters $w$ can be optimised along the time index  $t$ via online gradient descent, yielding the time-varying solutions
\begin{align}\label{theta_upd}
    \theta_{t+1}\leftarrow  \theta_{t} - \eta \nabla_\theta G_{t,\theta_t,w_t}
\end{align} and \begin{align}\label{w_upd}
    %w_{t+1} \leftarrow  w_{t} - \mu \nabla_w G_{t,\theta_t,w_t},
    w_{t+1} \leftarrow  w_{t} - \mu \nabla_w p_w (x=x_t | \mathbf{s}_t)
\end{align}
where $\eta>0$ and $\mu>0$ denote  learning rates.

% Finally, the scoring rule in \eqref{scoring_rule} requires access to the posterior belief ${p}_{\theta} (x \, | \, \mathbf{s})$. To this end, we use an estimator $E_w$, parameterised by $w$, that produces a fitted estimate $\hat{p}_{\theta, w} (x \, | \, \mathbf{s})$ of $p_{\theta} (x = \hat{x}_t \, | \, \mathbf{s})$, given the sequence of measurement outcomes $\mathbf{s}_t = [s_{0,t}, ..., s_{L-1, t}]$ with $s_{l,t} \sim p_{\theta}(s \, | \, x_t)$ for all $l \in [L]$.

% The optimisation is iterative: At time $t$, a new dataset $\mathcal{D}_t =\{\mathbf{s}_t, x_t\}$ is generated with the optimised probe and then the parameters of the estimator are optimised to fit the data. 

\subsection{Theoretical Guarantees}
Thanks to the properties of online conformal inference \cite{angelopoulos2024conformal}, the proposed dynamic optimisation strategy satisfies the target condition (\ref{barRo}). This is formalized in the following proposition.
\begin{proposition} \cite{angelopoulos2024conformal} 
For any sequence of target parameters, the long-term estimation loss achieved by the proposed scheme
can be upper  bounded as  
\begin{align}\label{LTrisk}
    \bar{L}(T) \leq \alpha + \frac{L_{\text{max}} + \max_{1 \leq t \leq T} \eta_{t}}{T} ||\Delta_{1:T}||,
\end{align}
where the sequence $||\Delta^{1:T}||$ takes the values
\begin{align}\label{Delta}
    ||\Delta_{1}|| = 1/\eta_{1} \,\,\, \text{and} \,\,\, ||\Delta_t|| = 1/\eta_{t} - 1/\eta_{t-1}.
\end{align}
\end{proposition}

\section{Experiments}
In this section, we provide experimental results to validate the proposed variational quantum sensing scheme. The code for the experiments is available at \cite{Git}.

\subsection{Tasks}
%We consider two tasks:
%\subsubsection{Magnetometry} 
A practical application of quantum sensing is magnetometry \cite{aiello2013composite, maayani2019distributed}. In this setting, the probe interacts with a uniform magnetic field. The parameter of interest $x_t$ represents a random phase,  which is acquired by the probe  after precessing for some time under the action of the magnetic field. The quantum channel is local and modeled as $\mathcal{E} (x_t) =  R_z (x_t)^{\otimes n}$, thus applying separately to each of the $n$ qubits of the probe. The phase is assumed to take values $x_t \in [0, \pi]$ and is discretised into $M=10$ equally spaced levels.

\subsection{Architecture and Hyperparameters}\label{sec:arch}
The quantum circuit used to generate the probe follows an $S(N)$-equivariant ansatz, which preserves permutation-equivariance \cite{schatzki2022theoretical}. The ansatz consists of general single qubit gates and of parametric two-qubit gates applied in a cyclical manner across all pairs of successive qubits. All single-qubit gates in the same layer share parameters, and, similarly, all two-qubit gates in the same layer share parameters.   The ansatz is comprised of $4$ layers and $n=4$ qubits. 

We measure local observables over $L=10$ probes for each time step $t$. For the optimisation of the variational parameters $\theta$ in \eqref{soft_X}, we use $\tau = 0.5$. The learning rate in \eqref{theta_upd} is set to $0.001$.

The estimator is an LSTM with two hidden layers with $1000$ hidden neurons each and ReLu activations. We use L2 regularisation and a multiplicative learning rate decay schedule in \eqref{w_upd}, with decay of $0.1$ every $50$ time steps. As further detailed below, we also study  Bayesian variants of the LSTM model based on ensembling \cite{pearce2018bayesian} and dropout \cite{gal2016dropout}. These models are known to be better able to capture the inherent epistemic uncertainty arising from the use of limited training data.  

 The parameters $\theta_1$ and $w_1$ in Algorithm 1 are initialized by pre-training the probe circuit and the estimator  using a small dataset $\mathcal{D}_t =\{\mathbf{s}_t, x_t\}_{t=1}^{20}$ of $20$ data samples.

Performance is measured using the coverage loss \eqref{hitt}, and thus the value $1-\alpha$ represents the target average coverage level. Results are averaged over five independent trials. 

%For the benchmark scheme, we use grid search to find the optimal hyperparameters.

\subsection{Benchmarks}\label{sec:bench}
As benchmarks, we consider the following schemes, all  using the same architectures  and hyperparameters for ansatz and  estimator as described in Sec.~\ref{sec:arch}:\begin{itemize}
\item \emph{Static}: This basic scheme fixes the threshold  $\lambda_t=\lambda$ in the  estimation set (\ref{set_predict}) by choosing it uniformly in the interval $[0,2]$, and it fixes the variational parameters of probes and estimator to respective their initializations. 
\item \emph{Static threshold}:  
To highlight the possible  benefits of using an online update rule, we also consider a  scheme that constructs the estimation set (\ref{set_predict}) with a fixed threshold $\lambda_t=\lambda$, selected as in the fully static scheme, while still updating probe and estimator parameters as in \eqref{theta_upd}. 
\item \emph{Static probe and estimator}: 
To test the performance of the dynamic probe and estimator optimisation strategy presented in \eqref{theta_upd} in isolation, we  finally 
 consider the case where probe and estimator are fixed to their initializations, while  the threshold $\lambda_t$ is adapted as  in \eqref{lambda_update}. 
\end{itemize}

\begin{figure}[tbp]
\centering
\includegraphics[width=0.45\textwidth]{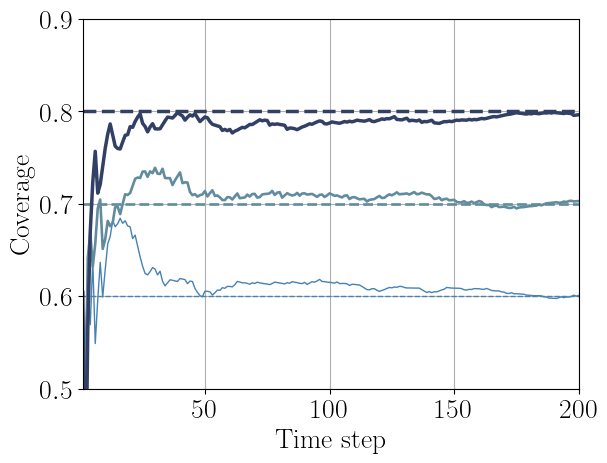}
\caption{Time-averaged coverage of the proposed dynamic scheme as a function of the time steps  $t$ for  target average loss values $\alpha=0.2$, $\alpha=0.3$ and $\alpha=0.4$.}
\label{fig:risk_epochs_map}
\end{figure}

\subsection{Results}
\begin{figure}[tbp]
\centering
\includegraphics[width=0.45\textwidth]{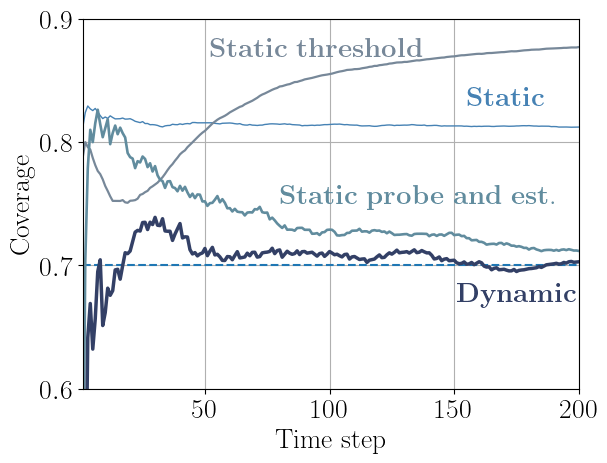}
\caption{Time-averaged coverage of the proposed dynamic scheme and of the benchmarks as a function of the time step $t$ for $\alpha=0.3$. The target coverage is shown in a dashed line.}
\label{fig:risk_epochs_map_bench}
\end{figure}

\begin{figure}[tbp]
\centering
\includegraphics[width=0.45\textwidth]{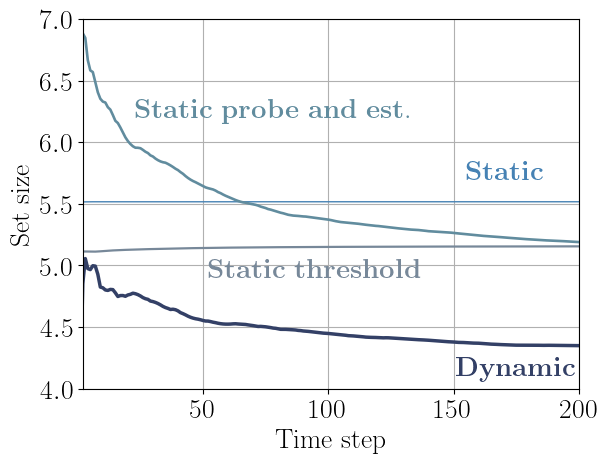}
\caption{Time-averaged set size of the proposed dynamic scheme and of the benchmarks as a function of the time step $t$ for $\alpha=0.3$.}
\label{fig:set_epochs_map}
\end{figure}

We first validate the capacity of the dynamic threshold adaptation strategy 
 \eqref{lambda_update} to  ensure the empirical coverage level $1-\alpha$ as the time step $t$ increases for  $\alpha=0.2$, $\alpha=0.3$, and $\alpha=0.4$. The time-averaged coverage $1-
    \bar{L}(t)$, where $\bar{L}(t)$ is defined in \eqref{barR}, is shown in Fig.~\ref{fig:risk_epochs_map}, confirming that the proposed scheme can achieve arbitrary coverage rates as stated in Proposition 1.

Then, we compare the performance of the proposed dynamic approach against the benchmarks described in Sec.~\ref{sec:arch}. We show the achieved coverage as a function of the number of epochs in Fig.~\ref{fig:risk_epochs_map_bench} and the time-averaged set size  in Fig.~\ref{fig:set_epochs_map} for $\alpha=0.3$. 

The results in Fig.~\ref{fig:risk_epochs_map_bench} confirm that the dynamic threshold adaptation \eqref{lambda_update} can achieve the desired coverage even with fixed probe and estimator  parameters. However, as seen in Fig.~\ref{fig:set_epochs_map}, the proposed dynamic optimization of variational parameters  obtains smaller, and thus more informative, sets. 

%Conversely, using the given fixed threshold results in including all values in the constructed set, yielding perfect coverage but an uniformative estimate.

\begin{figure}[tbp]
\centering
\includegraphics[width=0.45\textwidth]{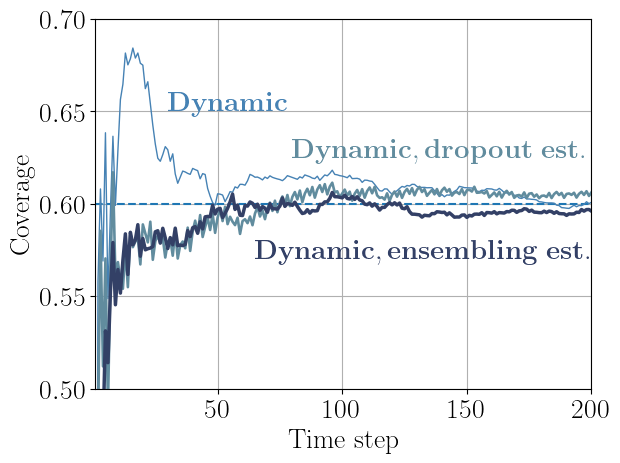}
\caption{Time-averaged coverage as a function of the number of time steps $t$ for the proposed dynamic scheme with different decoder models.}
\label{fig:risk_epochs_map_ensdrop}
\end{figure}

%As \eqref{scoring_rule} shows, the estimation list critically depends on the quality of the model that produces $\hat{p}_{\theta, \mu, w} (x \, | \, \mathbf{s})$. To improve $\hat{p}_{\theta, \mu, w} (x \, | \, \mathbf{s})$,

Finally, we demonstrate the importance of accounting for epistemic uncertainty  at the estimator side by evaluating the performance of Bayesian 
 learning techniques, namely ensembling \cite{pearce2018bayesian} and dropout \cite{gal2016dropout}. The former uses an ensemble of $5$ LSTMs to make predictions, whilst the latter drops a portion,set to 0.4, of the neurons in each forward pass when making a prediction whereby repeated passes yield distinct predictions. In Fig.~\ref{fig:risk_epochs_map_ensdrop}, we show the coverage as a function of the number of probes, for a target loss of $\alpha=0.4$. As can be seen, both methods improve the set construction and meet the target coverage requirement.

%The estimation loss in \eqref{est_loss}, with the distance measure defined as $$\ell(x, s) = \frac{1}{T}\sum_{t=1}^{T-1}|\hat{x}_t - x_t|$$ is shown in Fig.~\ref{fig:testing_error}. With this definition, the estimation loss practically represents the bias of our estimator. 

%The bias of the estimator is shown in Fig.~\ref{fig:testing_error}. As Fig.~\ref{fig:testing_error} shows, the performance of the benchmark improves as the number of probes increases. This is expected since GHZ states are known to be optimal in asymptotic regimes. As the number of probes increases, the prediction loss decreases, and the estimator becomes less biased. Similarly, list based optimisation appears to match and at times perform better than using the Fisher Information as an objective. Unlike the Fisher Information, however, it is immune to barren plateaus \cite{maclellan2024end}.

\section{Conclusions}

We have presented an estimation loss control framework for variational quantum sensing. The proposed method can theoretically quarantee the long-term estimation loss. Experimental results confirm that it can effectively control the estimation loss, whilst providing useful estimates. 

Various future research directions arise. For example, combining the proposed method with explicit noise mitigation techniques might be able to help further in parameter estimation. More sophisticated loss construction rules might also be beneficial. In this regard, it would also be beneficial to theoretically characterise the length of the set, as it ultimately quantifies the estimator uncertainty \cite{zecchin2024generalization}.

\section*{Acknowledgments}
The work of H. Joudeh was supported by the European Research Council (ERC) project IT-JCAS (101116550).
%%%%%
The work of R. van Sloun was supported by the European Research Council (ERC) project US-ACT (101077368) 
%under the ERC starting grant nr. 101077368 (US-ACT). 
The work of O. Simeone was supported by the European Union’s Horizon Europe project CENTRIC (101096379), by an Open Fellowship of the EPSRC (EP/W024101/1), and by the EPSRC project (EP/X011852/1).

\bibliographystyle{IEEEtran}
\bibliography{litdab.bib}

\end{document}